\documentclass[11pt]{article}
\usepackage{amsfonts,amsmath}
\usepackage{graphicx}

\def\bbt{\bibitem}
\def\be{\begin{equation}}
\def\en{\end{equation}}
\def\ber{\begin{eqnarray}}
\def\enr{\end{eqnarray}}
\def\nmb{ \nonumber\\}
\def\d{\partial}
\def\rbr{\rbrack}
\def\lbr{\lbrack}
\def\rbrc{\rbrace}
\def\lbrc{\lbrace}

\def\ov{\over }

\def\sgm{\sigma}

\def\al{\alpha}
\def\bet{\beta}
\def\gm{\gamma}
\def\Gm{\Gamma}
\def\im{\imath}

\def\lm{\lambda}

\def\om{\omega}
\def\et{\eta}
\def\tt{\theta}
\def\Tt{\Theta}

\def\dlt{\delta}

\def\vphi{\varphi}
\def\cz{{\cal{Z}}}
\def\cw{{\cal{W}}}

\begin{document}


\vskip 2 true cm
\centerline{\bf GENERALIZED K$\ddot{A}$HLER GEOMETRY IN KAZAMA-SUZUKI COSET MODELS.}
\vskip 20pt 
\centerline{S. E. Parkhomenko}
\vskip 20pt
\centerline{L.D.Landau Institute for Theoretical Physics}
\centerline{Akademika Semenova av. 1-A}
\centerline{Chernogolovka, 142432 Moscow region, Russia.}

\vskip 1 true cm
\centerline{\bf Abstract}
\vskip 0.5 true cm

It is shown that Kazama-Suzuki conditions for the denominator subgroup of N=2 superconformal $G/H$ coset model determine Generalized K$\ddot{a}$hler geometry on the target space of the corresponding N=2 supersymmetric $\sgm$-model.

\smallskip
\vskip 10pt
\leftline{\bf 1.Introduction.}

 It is by now well-known due to Gepner \cite{Gep} that the unitary N=2 superconformal field theories play an important role in the construction of realistic models of superstring compactification from 10 to 4 dimensions. Gepner's idea is to use N=2 superconformal field theories with central charge 9 for the internal sector of the string degrees of freedom and apply the GSO projection in such a way to be consistent with modular invariance. In particular, Gepner considered a product of N=2 minimal models such that their total central charges adds up to 9 and shown close relationship of his purely algebraic construction to the geometric Calabi-Yau $\sgm$-models compactification. 

 The Calabi-Yau manifold being a complex K$\ddot{a}$hler manifold is not accidental but caused by a close connection between the extended supersymmetry and K$\ddot{a}$hler geometry. In a more general case the background geometry may include also an antisymmetric $B$-field. In that case the corresponding 2-dimensional supersymmetric $\sgm$-model have a second supersymmetry when the target-space has a bi-Hermitian geometry, known also as Gates-Hull-Ro$\check{c}$ek geometry \cite{GHR}. In this situation the target manifold contains two complex structures with a Hermitian metric with respect to each of the complex structures. Quite recently it has been shown in \cite{Gualt} that these set of geometric objects, metric, antisymmetric $B$-field and two complex structures antisymmetric with respect to the metric have a unified description in the context of Generalized K$\ddot{a}$hler (GK) geometry. It allowed to develop for these models the GK geometry construction of $N=2$ Virasoro superalgbra \cite{Z}, \cite{BLPZ}, \cite{Zlec}, \cite{HelZ}, \cite{HelZ1}. It make sense by this reason to investigate the relationship of GK geometry $\sgm$-models to $N=2$ superconformal field theories more closely. 

 The $N=2$ supersymmetric WZW models on the compact groups \cite{SSTP}, \cite{P} provide a large class of examples where this relation has been studied quite well \cite{SevT}, \cite{RSSev}, 
\cite{SevSW}, \cite{SevST}, \cite{PGKG}. These are the exactly solvable quantum conformal $\sgm$-models whose targets supports simultaneously GK geometry causing the extended $N=2$-supersymmetry and affine Kac-Moody superalgebra structure ensuring the exact solution. 

 In a more general context it would be important to see what other unitary $N=2$ superconformal field theories can be related to the GK geometry $\sgm$-models.  A large class of of unitary N=2 superconformal field theories, including N=2 superconformal minimal models, has been constructed by Kazama and Suzuki \cite{KS} using the coset space method \cite{GKO}.

 In this note we consider their construction from the point of view of GK geometry. We show that Kazama-Suzuki conditions for the denominator subgroup $H$ of N=2 superconformal $G/H$ coset model determine in classical limit GK geometry on the target space of the corresponding $\sgm$-model. To do that we use $N=(1,1)$ superspace and reformulate in Section 2 the Kazama-Suzuki conditions for the denominator subgroup $H$ in terms of Manin triples. For further analysis we use Hamiltonian formalism which has already been successfuly applied for the studying of $N=2$ superconformal $\sgm$-models GK geometry (see for example \cite{Z}, \cite{Zlec}, \cite{BLPZ}). In particular, we represent the Hamiltonian formulation of $N=2$ supersymmetric WZW model in Section 3 and relate the bi-Hermitian geometry of the model to the bi-Poisson geometry of 
S.Lyakhovich and M.Zabzine \cite{LyakhZ} analysing $N=(2,2)$ Virasoro superalgebra in the classical limit. In Section 4 we generalize the disscussion of Section 3 to the case of Kazama-Suzuki models using Manin triples language from Section 2.
Then we show that Kazama-Suzuki conditions for the denominator subgroup $H$ determine bi-Poisson geometry
on the target space of the corresponding $\sgm$-model which comes from a pair of covariantly constant target-space complex structures which are skew-symmetric w.r.t to the metric. It allows us to establish GK geometry of the target spaces of Kazama-Suzuki models. In Section 5 we summarize the results and disscuss future directions to investigate.

\vskip 10 pt
\leftline{\bf 2. Manin triple construction of Kazama-Suzuki coset model.}


 I start with some preliminaries about $N=2$ superconformal WZW model on the compact group $G$ and recall breafly the Manin triple construction of Kazama-Suzuki coset model represented in \cite{PKS} (see also \cite{HS}).

 In this case the group manifold $G$ is even dimensional and endowed with right-invariant complex structure $J_{L}$ and left-invarint complex structure $J_{R}$. Both of them are skew-symmetric w.r.t. the invariant metric on the group. Identifying the Lie algebra $g$ of the group with left-invariant vector fields or right-invariant vector fields we obtain the complex structure $J$ on $g$. It endows the complexification 
$g^{\mathbb{C}}$ of $g$ with the Manin triple structure \cite{Drin} which is the triple $(g^{\mathbb{C}},g_{+},g_{-})$ consisting of a Lie algebra $g^{\mathbb{C}}$, with nondegenerate invariant inner product
$<,>$ and isotropic Lie subalgebras $g_{\pm}$ such that
$g^{\mathbb{C}}=g_{+}\oplus g_{-}$ as a vector space. It is clear that the subalgebras $g_{\pm}$ are $\pm\im$-eigenspaces of the complex structure and the real Lie algebra $g$ is given by the fixed point set w.r.t. the natural antilinear involution which conjugates isotropic subalgebras,
$\tau: g_{\pm}\to g_{\mp}$. It is not difficult to establish a
correspondence between the complex Manin triples endowed with antilinear involution $\tau$ conjugating the isotrophic subalgebras and complex structures on the real Lie algebras \cite{P}. Due to this correspondence Manin triple construction of Kazama-Suzuki models presented in \cite{PKS} can be connected to the approach of \cite{HS} and \cite{SSTP}based on the complex structures on Lie algebras.

 Let us fix arbitrary
orthonormal basis $\{E^{A}, E_{A}$, $A= 1,...,{1\ov 2}dim g\}$ in algebra
$g^{\mathbb{C}}$ so that $\{E^{A}\}$ is a basis in $g_{-}$, $\{E_{A}\}$
is a basis in $g_{+}$. In this basis the commutators 
have the form
\ber  \lbr E^{A},E^{B}\rbr=f^{AB}_{C}E^{C},     \
           \lbr E_{A},E_{B}\rbr=f_{AB}^{C}E_{C},     \nmb
           \lbr E^{A},E_{B}\rbr=f_{BC}^{A}E^{C}-f^{AC}_{B}E_{C}, 
\label{1.Mtrip1}
\enr

On the two-dimensional superspace with holomorphic supercoordinates $\cz=(z, \Tt)$ we use $L^{A}(\cz)$, $L_{A}(\cz)$ to denote holomorphic (left-moving) spin-$1/2$ super-currents valued in $g_{-}$ and $g_{+}$ correspondingly.  Similarly we denote by $R^{A}(\bar{\cz})$, $R_{A}(\bar{\cz})$ the anti-holomorphic (right-moving) spin-$1\ov 2$ supercurrents valued in $g_{-}$ and $g_{+}$. 

 The holomorphic currents satisfy the OPE's
\ber
L^{A}(\cz_{1})L^{B}(\cz_{2})=(\cz_{1}-\cz_{2})^{-{1\ov 2}}f^{AB}_{C}L^{C}(\cz_{2})+...
\nmb
L_{A}(\cz_{1})L_{B}(\cz_{2})=(\cz_{1}-\cz_{2})^{-{1\ov 2}}f_{AB}^{C}L_{C}(\cz_{2})+...
\nmb
L^{A}(\cz_{1})L_{B}(\cz_{2})=(\cz_{1}-\cz_{2})^{-1}k\dlt^{A}_{B}+
(\cz_{1}-\cz_{2})^{-{1\ov 2}}(f_{BC}^{A}L^{C}-f^{AC}_{B}L_{C})(\cz_{2})+...
\label{1.Lcurrents}
\enr
where $\cz_{1}-\cz_{2}=z_{1}-z_{2}-\Tt_{1}\Tt_{2}$ and $(\cz_{1}-\cz_{2})^{-{1\ov 2}}=(\cz_{1}-\cz_{2})^{-1}(\Tt_{1}-\Tt_{2})$. The currents  $R^{A}(\bar{\cz})$, $R_{A}(\bar{\cz})$ satisfy similar OPE's. In what follows we concentrate on the holomorphic sector of the model.

 Having the Manin triple structure and the OPE's above we can construct \cite{P} the spin-1 supercurrent of the $N=2$ Virasoro superalgebra
\ber
K_{g}(\cz)={1\ov k}(:L^{A}L_{A}:+f_{A}DL^{A}-f^{A}DL_{A})
\label{1.K}
\enr
where $f_{A}=f_{AB}^{B}$, $f^{A}=f^{AB}_{B}$. This current generates stress-energy spin-3/2 supercurrent of $N=2$ Virasoro superalgebra 
\ber
\Gm_{g}(\cz)={1\ov k}(:DL^{A}L_{A}:+:DL_{A}L^{A}:+{1\ov k}f_{AB}^{C}:L_{C}:L^{A}L^{B}::+{1\ov k}f^{AB}_{C}:L^{C}:L_{A}L_{B}::)
\label{1.SEnergy}
\enr
by the OPE
\ber
K_{g}(\cz_{1})K_{g}(\cz_{2})=(\cz_{1}-\cz_{2})^{-2}c_{g}+(\cz_{1}-\cz_{2})^{-{1\ov 2}}\Gm_{g}(\cz_{2})+...
\label{1.KKope}
\enr
where 
\ber
c_{g}=3({1\ov 2}dim g+{2\ov k}f^{A}f_{A})
\label{1.c}
\enr

 Let us fix some Manin subtriple which is invariant under the involution $\tau$
\ber
(h^{\mathbb{C}}\subset g^{\mathbb{C}}, h_{+}\subset g_{+}, h_{-}\subset g_{+})
\label{1.Msubtrip}
\enr
In this case the direct sum $h_{+}\oplus h_{-}$ is a complexification $h^{\mathbb{C}}$ of a real subalgebra $h$.
$L^{\al}(\cz)$ is used to denote the currents valued in $h_{-}$, $L_{\al}(\cz)$ is used to denote the currents valued in
$h_{+}$. Because of the Manin subtriple is fixed one can construct corresponding spin-1 supercurrent $K_{h}$ 
\ber
K_{h}(\cz)={1\ov k}(:L^{\al}L_{\al}:+\vphi_{\al}DL^{\al}-\vphi^{\al}DL_{\al})
\label{1.Kh}
\enr
where $\vphi_{\al}=f_{\al\bet}^{\bet}$, $\vphi^{\al}=f^{\al\bet}_{\bet}$. $K_{h}(\cz)$ generates super stress-energy tensor of 
the $N=2$ WZW model associated to the denominator subgroup $H$ of the coset model. 

 In the papers \cite{KS} Kazama and Suzuki found the conditions the denominator subgroup $H$ must satisfy in order to $N=1$ superconformal $G/H$ coset model be $N=2$ superconformal. Their conditions were eqivalently reformulated in terms of the Manin triple and Manin subtriple in \cite{PKS}: 

{\bf Proposition.}

 The $N=1$ superconformal coset model $G/H$ is $N=2$ superconformal if the subspaces $t_{\pm}=g_{\pm}\backslash h_{\pm}$ are subalgebras. 
In this case the holomorphic spin-1 supercurrent
\ber
K_{KS}=K_{g}-K_{h}
\label{1.KKS}
\enr
satisfy Kazama-Suzuki conditions: 
\ber
L^{\al}(\cz_{1})K_{KS}(\cz_{2})=reg., \ L_{\al}(\cz_{1})K_{KS}(\cz_{2})=reg.
\label{1.KSconditions}
\enr
and generates $N=2$ Virasoro superalgebra of the coset-model with super stress-energy tensor
\ber
\Gm_{KS}=\Gm_{g}-\Gm_{h}
\label{1.cosetenergy}
\enr
and central charge $c_{KS}=c_{g}-c_{h}$. The similar statement is true also in anti-holomorphic sector of the model.

 In \cite{CrRoScho} this construction was generalized to the supergroup manifolds. In \cite{Ahn} some particular examples of this general construction was considered to build the W-algebra currents.


 In \cite{PKS} the proposition was proved in component fields. Here I sketch the proof of the proposition using $N=1$ superfield formalism which will be crucially for a geometric interpretation of Kazama-Suzuki construction. 

 Let $L^{a}(\cz)$ be the $t_{-}$-valued basic currents and $L_{a}(\cz)$ be the $t_{+}$-valued basic currents.
The calculation of the first OPE from (\ref{1.KSconditions}) gives the result:
\ber
kL^{\al}(\cz_{1})K_{KS}(\cz_{2})=-(\cz_{1}-\cz_{2})^{-1}<\lbr v-w, E^{\al}\rbr,E^{A}L_{A}(\cz_{2})+E_{A}L^{A}(\cz_{2})>+
\nmb
(\cz_{1}-\cz_{2})^{-{1\ov 2}}(<\lbr E_{a},E_{b}\rbr_{J},E^{\al}>:L^{a}L^{b}:+
\nmb
2<\lbr E_{a},E^{b}\rbr_{J},E^{\al}>:L^{a}L_{b}:+
\nmb
<\lbr E^{a},E^{b}\rbr_{J},E^{\al}>:L_{a}L_{b}:)(\cz_{2})+reg.
\label{1.LhKope}
\enr
where $v={1\ov k}(f^{A}E_{A}-f_{A}E^{A})$, $w={1\ov k}(\vphi^{\al}E_{\al}-\varphi_{\al}E^{\al})$ and $\lbr , \rbr_{J}$ is a new Lie algebra bracket on vector space $g^{\mathbb{C}}$: for any vectors $x,y\in g^{\mathbb{C}}$
\ber
\lbr x, y\rbr_{J}={1\ov 2}(\lbr Jx,y\rbr+\lbr x, Jy\rbr)
\label{1.Jbrack}
\enr
It is easy to see that 
\ber
\lbr g_{+},g_{-}\rbr_{J}=0,
\nmb
\lbr g_{+},g_{+}\rbr_{J}=\im\lbr g_{+},g_{+}\rbr,
\nmb 
\lbr g_{-},g_{-}\rbr_{J}=-\im\lbr g_{-},g_{-}\rbr
\label{1.Jbrack1}
\enr 
Because of $t_{\pm}$ are isotropic subalgebras one can see that they are ideals in $g_{\pm}$. Therefore 
\ber
<\lbr E_{a},E_{b}\rbr_{J},E^{\al}>=0
\label{1.Jbrack2}
\enr
as it follows from (\ref{1.Jbrack1}). It follows also from (\ref{1.Jbrack1}) that 
\ber
<\lbr E^{a},E^{b}\rbr_{J},E^{\al}>=0
\label{1.Jbrack3}
\enr
Thus, $(\cz_{1}-\cz_{2})^{-{1\ov 2}}$ contribution from (\ref{1.LhKope}) vanishes.

 Consider the $(\cz_{1}-\cz_{2})^{-1}$ contribution (it is absent in the classical limit). We find
\ber
<\lbr v-w, E^{\al}\rbr, E^{\al}\rbr,E^{A}L_{A}(\cz_{2})+E_{A}L^{A}(\cz_{2})>=
\nmb
<f_{A}(f^{A\al}_{b}E^{b}+f^{A\al}_{\gm}E^{\gm})-\vphi_{\bet}f^{\bet\al}_{\gm}E^{\gm},E^{\al}\rbr,E^{A}L_{A}(\cz_{2})+
E_{A}L^{A}(\cz_{2})>+
\nmb
<f^{A}(f_{Ac}^{\al}E^{c}+f_{A\gm}^{\al}E^{\gm}-f^{\al b}_{A}E_{b}-f^{\al\gm}_{A}E_{\gm})-
\vphi^{\bet}(f_{\bet\gm}^{\al}-f^{\al\gm}_{\bet}E_{\gm}),E^{\al}\rbr,E^{A}L_{A}(\cz_{2})+E_{A}L^{A}(\cz_{2})>
\label{1.LhKope1}
\enr
Because of $h_{\pm}$ are subalgebras and $t_{\pm}$ are ideals in $g_{\pm}$ some of the structure constants in the expression 
(\ref{1.LhKope}) are zero so that we obtain
\ber
 k<\lbr v-w, E^{\al}\rbr, E^{\al}\rbr,E^{A}L_{A}(\cz_{2})+E_{A}L^{A}(\cz_{2})>=
\nmb
<\chi_{a}f^{a\al}_{b}E^{b}+\chi_{\bet}f^{\bet\al}_{\gm}E^{\gm},E^{\al}\rbr,E^{A}L_{A}(\cz_{2})+E_{A}L^{A}(\cz_{2})>+
\nmb
<-\chi^{a}f^{\al b}_{a}E_{b},E^{A}L_{A}(\cz_{2})+E_{A}L^{A}(\cz_{2})>+
\nmb
<f^{\bet}(f_{\bet c}^{\al}E^{c}+f_{\bet\gm}^{\al}E^{\gm}-f^{\al\gm}_{\bet}E_{\gm}),E^{\al}\rbr,E^{A}L_{A}(\cz_{2})+
E_{A}L^{A}(\cz_{2})>-
\nmb
<\phi^{\bet}(f_{\bet\gm}^{\al}E^{\gm}-f^{\al\gm}_{\bet}E_{\gm}),E^{\al}\rbr,E^{A}L_{A}(\cz_{2})+E_{A}L^{A}(\cz_{2})>
\label{1.LhKope1}
\enr
where
\ber
\chi_{A}=f_{Ab}^{b}, \ \chi^{A}=f^{Ab}_{b}
\label{1.chi}
\enr
The relations 
\ber
\chi_{A}f_{aB}^{A}=0, \ \chi^{A}f^{aB}_{A}=0
\nmb
\chi_{A}f^{\mu A}_{a}=0, \ \chi^{A}f_{\mu A}^{a}=0
\label{1.chirelations}
\enr
which follow from the Jacobi identity for $g^{\mathbb{C}}$ and because of $t_{\pm}$ are ideals, reduce the expression above to 
\ber
 k<\lbr v-w, E^{\al}\rbr, E^{A}L_{A}(\cz_{2})+E_{A}L^{A}(\cz_{2})>=
\nmb
<\chi_{\bet}f^{\bet\al}_{\gm}E^{\gm}+\chi^{\bet}f_{\bet\gm}^{\al}E^{\gm},E^{A}L_{A}(\cz_{2})+E_{A}L^{A}(\cz_{2})>
\label{1.LhKope2}
\enr
But it is zero because of
\ber
\chi_{\bet}f^{\bet\al}_{\gm}+\chi^{\bet}f_{\bet\gm}^{\al}=0
\label{1.chirealtions1}
\enr
It can be proven in turn as follows: due to the Jacoby identity for $g^{\mathbb{C}}$ we have the relation
\ber
f_{C}f^{CB}_{A}+f^{C}f_{CA}^{B}=f_{DC}^{B}f^{DC}_{A},  
\label{1.Jacconseq}
\enr
Specifying this for $A=\al$, $B=\bet$ we obtain
\ber
f_{C}f^{C\bet}_{\al}+f^{C}f_{C\al}^{\bet}=f_{DC}^{\bet}f^{DC}_{\al},  
\label{1.Jacconseq1}
\enr
because of $t_{\pm}$ are the ideals this relation is equivalent to
\ber
f_{\gm}f^{\gm\bet}_{\al}+f^{\gm}f_{\gm\al}^{\bet}=(\chi+\vphi)_{\gm}f^{\gm\bet}_{\al}+(\chi+\vphi)^{\gm}f_{\gm\al}^{\bet}=
f_{\nu\mu}^{\bet}f^{\nu\mu}_{\al},  
\label{1.Jacconseq2}
\enr
but $h_{\pm}$ are isotropic subalgebras of the Manin triple $(h, h_{+},h_{-})$ so that
\ber
\vphi_{\gm}f^{\gm\bet}_{\al}+\vphi^{\gm}f_{\gm\al}^{\bet}=f_{\nu\mu}^{\bet}f^{\nu\mu}_{\al},  
\label{1.Jacconseqh}
\enr
which proves (\ref{1.chirealtions1}). 

 Hence, $(\cz_{1}-\cz_{2})^{-1}$ contribution is zero also, so the first OPE from (\ref{1.KSconditions}) is correct. Analogously the second OPE from (\ref{1.KSconditions}) can be proved. 

 Similar Lie algebra analysis can be performed to establish the OPE
\ber
K_{KS}(\cz_{1})K_{KS}(\cz_{2})=(\cz_{1}-\cz_{2})^{-2}(c_{g}-c_{h})+(\cz_{1}-\cz_{2})^{-{1\ov 2}}(\Gm_{g}-\Gm_{h})(\cz_{2})+...
\label{1.KksKks}
\enr
as well as the other $N=2$ Virasoro superalgebra OPE's.

\leftline{\bf Remark.} The Proposition can be reformulated in a more economical way. Using the invariant metric one can identify the $g^{\mathbb{C}}$ and $h^{\mathbb{C}}$ with the dual spaces $(g^{\mathbb{C}})^{*}$ and $(h^{\mathbb{C}})^{*}$. We obtain thereby two bialgebras \cite{SemTian} $(g^{\mathbb{C}},(g^{\mathbb{C}})^{*})$ and $(h^{\mathbb{C}},(h^{\mathbb{C}})^{*})$, where the Lie algebra structure on the dual spaces is given by the brackets (\ref{1.Jbrack}). Then the conditions the subspaces $t_{\pm}$ must satisfy are equivalent to the statement that $(h^{\mathbb{C}},(h^{\mathbb{C}})^{*})$ is a subbialgebra in $(g^{\mathbb{C}},(g^{\mathbb{C}})^{*})$ 
\cite{SemTian}.

\vskip 10 pt
\leftline{\bf 3.Hamiltonian formulation of classical $N=2$ supersymmetric gauged WZW model}
\leftline{\bf and bi-Poisson structure.}

 Here we breafly discuss the Hamiltonian formalism in the classical $N=2$ supersymmetric WZW model and provide bi-Poisson definition of GK geometry. 

 This formalism was considered in \cite{PGKG}, \cite{HelZ}, \cite{HelZ1} (N=0,1 versions was considered in \cite{Fedya}, see also \cite{Bolg}). There it was shown that phase space of the model is a sheaf of twisted Poisson Vertex algebras \cite{Hel}, \cite{Bryl}, \cite{EHKZ}.
Locally, the sections of the sheaf are generated by the canonically conjugated $N=1$ superfields $X^{\mu}(Z)$, $X^{*}_{\mu}(Z)$ with the canonical Poisson super-brackets
\ber
\lbrc X^{*}_{\mu}(Z_{1}), X^{\nu}(Z_{2})\rbrc=-\lbrc X^{\nu}(Z_{2}), X^{*}_{\mu}(Z_{1})\rbrc=
\dlt^{\nu}_{\mu}\dlt(Z_{1}-Z_{2})
\label{2.PBcanon}
\enr
and theirs super-derivatives along the super-circle variable $Z=(\sgm,\tt)$. The spin-0 superfields $X^{\mu}(Z)$ (generalized coordinates) come from the local coordinates $x^{\mu}$ on the group manifold $G$ while the conjugated spin-${1\ov 2}$ superfields $X^{*}_{\mu}(Z)$ (generalized momenta) correspond to the derivatives ${\d\ov \d x^{\mu}}$ (in physics literature this set of fields also known as a $b-c-\bet-\gm$ system). The super-Hamiltonian of the model is given by a zero mode of the corresponding stress-energy supertensor component. 

 On the intersection of patches with local coordinates $x^{\mu}$ and $y^{\mu}$ the canonical superfields are related by
\ber
Y^{\nu}(Z)=y^{\nu}(X^{\mu}(Z)),
\nmb
Y^{*}_{\nu}={\d x^{\mu}\ov \d y^{\nu}}(X^{*}_{\mu}+(B^{y}-B^{x})_{\mu\lm}DX^{\lm})
\label{2.momentrans}
\enr
The set of 2-forms $B^{x}$, $B^{y}$, ..., defines the set of 1-forms $A^{yx}$, ..., determined on the intersections such that
\ber
dB^{x}=dB^{y}=\cal{H}, 
\nmb
B^{y}-B^{x}=dA^{yx}
\label{2.gerbe1}
\enr
where $\cal{H}$ is a 3-form on $G$ specifying the level $k$ of WZW model. When $k\in \mathbb{Z}$ this data, toghether with the complex structures $J_{L,R}$, define biholomorphic gerbe with connection \cite{Bryl}, \cite{HLRUZ}. One can see that we are in the situation of theorem 
from paper \cite{HLRUZ} where very nice description of GK geometry in terms of locally defined symplectic forms determining the biholomorphic gerbe was found. As a result we get the global description of phase space of $N=2$ supersymmetric WZW model \cite{PGKG}. 

 In terms of the local coordinates the exact construction of the Kac-Moody superalgebra currents in $N=2$ WZW model was given in \cite{PGKG}. I will not reproduce the expressions for the currents here but notice only that they are given by certain super-circle functions valued in the direct sum of tangent and cotangent bundle of the group manifold $G$. The Poisson brackets for the currents are given by
\ber
\lbrc L^{i}(Z_{1}), L^{j}(Z_{2})\rbrc=
k\et^{ij}\dlt'(Z_{1}-Z_{2})+
\dlt(Z_{1}-Z_{2})f^{ij}_{k}L^{k}(Z_{2})
\nmb
\lbrc R^{i}(Z_{1}), R^{j}(Z_{2})\rbrc=
-k\et^{ij}\dlt'(Z_{1}-Z_{2})-
\dlt(Z_{1}-Z_{2})f^{ij}_{k}R^{k}(Z_{2})
\nmb
\lbrc L^{i}(Z_{1}), R^{j}(Z_{2})\rbrc=0
\label{2.KMPB}
\enr
where $\dlt'(Z)$ means superderivative of $\dlt(Z)$, 
$L^{i}(Z)=(L^{A}(Z),L_{A}(Z))$,  $R^{i}(Z)=(R^{A}(Z),R_{A}(Z))$ and $\eta^{ij}$, $f^{ij}_{k}$ are the invariant metric componets and structure constants on Lie algebra $g$.

 Another advantage of Hamiltonian formalism is the identification of primary Kac-Moody superalgebra fields with the functions on $G$ so that the Poisson brackets characterizing  
the primary field $\Phi(Z)$
\ber
\lbrc L^{i}(Z_{1}),\Phi (Z_{2})\rbrc=
\dlt(Z_{1}-Z_{2})L^{i}\cdot\Phi(Z_{2})
\nmb
\lbrc R^{i}(Z_{1}),\Phi (Z_{2})\rbrc=
\dlt(Z_{1}-Z_{2})R^{i}\cdot\Phi(Z_{2})
\label{2.primary}
\enr
are given by the actions $L^{i}\cdot\Phi$ of the basic left translations (basic right-invariant vector fields on $G$) and by the actions 
$R^{i}\cdot\Phi$ of the basic right translations (basic left-invariant vector fields on $G$) on the function $\Phi$ determined on $G$. Thus, using the Hamiltonian formalism we obtain supersymmetric generalization \cite{Fedya} of the regular representation of Kac-Moody algebra \cite{FeP}, \cite{Fr}.
 
 Notice that two copies of $N=2$ Virasoro algebra are acting in the regular representation by the classical version of the quantum $N=2$ Sugawara construction from section 2. In particular, the spin-1 currents 
\ber
K^{L}(Z)={\im\ov 2k}\om_{Lij}L^{i}L^{j}(Z)={1\ov k} L^{A}L_{A}(Z), \
K^{R}(Z)=-{\im\ov 2k} \om_{Rij}R^{i}R^{j}(Z)=-{1\ov k} R^{A}R_{A}(Z)
\label{2.KScarrent}
\enr
can be used to determine skew-symmetric bilinear operations (brackets) on the primary fields $\Phi_{1,2}(Z)$ of the $N=2$ superconformal WZW model:
\ber
\lbrc \Phi_{1}(Z_{1}),\Phi_{2}(Z_{2})\rbrc_{K^{L}}:=\lbrc K^{L}_{-1/2}\Phi_{1}(Z_{1}),\Phi_{2}(Z_{2})\rbrc,
\nmb
\lbrc \Phi_{1}(Z_{1}),\Phi_{2}(Z_{2})\rbrc_{K^{R}}:=\lbrc K^{R}_{-1/2}\Phi_{1}(Z_{1}),\Phi_{2}(Z_{2})\rbrc
\label{2.Derbrack}
\enr
where
\ber
K^{L,R}_{-1/2}\Phi(Z)=\oint dW\lbrc K^{L,R}(W),\Phi(Z)\rbrc
\label{2.Derbrack1}
\enr
and $\om_{Lij}=\et_{ik}(J_{L})^{k}_{j}$, $\om_{Rij}=\et_{ik}(J_{R})^{k}_{j}$. 
Because of the primary fields correspond to the functions $\Phi_{1,2}$ on $G$, the residues of the brackets above define the skew-symmetric brackets on functions:
\ber
\lbrc \Phi_{1}(Z_{1}),\Phi_{2}(Z_{2})\rbrc_{K^{L}}=
\dlt(Z_{1}-Z_{2}){\im\ov 2k}\om_{Lij}(L^{i}\cdot\Phi_{1})(L^{j}\cdot\Phi_{2})(Z_{2})
\nmb
\lbrc \Phi_{1}(Z_{1}),\Phi_{2}(Z_{2})\rbrc_{K^{R}}=
-\dlt(Z_{1}-Z_{2}){\im\ov 2k}\om_{Rij}(R^{i}\cdot\Phi_{1})(R^{j}\cdot\Phi_{2})(Z_{2})
\label{2.Derbrack2}
\enr
The brackets (\ref{2.Derbrack}) do not satisfy Jacobi super-identity, instead the combinations
\ber
\lbrc \Phi_{1}(Z_{1}),\Phi_{2}(Z_{2})\rbrc_{\pm}:=
\lbrc \Phi_{1}(Z_{1}),\Phi_{2}(Z_{2})\rbrc_{K^{L}}\pm \lbrc \Phi_{1}(Z_{1}),\Phi_{2}(Z_{2})\rbrc_{K^{R}}
\label{2.Derbrack3}
\enr
do:
\ber
\lbrc \Phi_{1}(Z_{1}),\lbrc\Phi_{2}(Z_{2}),\Phi_{3}(Z_{3})\rbrc_{K^{L}\pm K^{R}}\rbrc_{\pm}-
\nmb
\lbrc \lbrc \Phi_{1}(Z_{1}),\Phi_{2}(Z_{2})\rbrc_{K^{L}\pm K^{R}},\Phi_{3}(Z_{3})\rbrc_{\pm}+
\lbrc \Phi_{2}(Z_{2}),\lbrc \Phi_{1}(Z_{1}),\Phi_{3}(Z_{3})\rbrc_{K^{L}\pm K^{R}}\rbrc_{\pm}=0
\label{2.Derbrack4}
\enr
This can be checked by the direct calculation if one takes into account that the structure constants for the right tralslations are opposite to the structure constants for left translations and use the integrability property of the complex structures $J_{L}$, $J_{R}$. 

 The residues of the brackets (\ref{2.Derbrack3}) define the pair of Poisson brackets on $G$. Looking at the classical version of (\ref{1.SEnergy}), (\ref{1.KKope}) one can see that these two brackets are not compatible but the Schouten bracket of the bivector associated to $K^{L}$ with the bivector associated to $K^{R}$ is proportional to the WZW 3-form $\cal{H}$ on the group $G$. In this way we have just came to the definition of Generalized K$\ddot{a}$hler geometry by the bi-Poisson structure found in \cite{LyakhZ}. Indeed, as we have seen, the spin-1 currents (\ref{2.KScarrent}) of the $N=(2,2)$ Virasoro superalgebra are given by the complex structures $J_{L,R}$ which are skew-symmetric with respect to the metric on group manifold $G$. Because of the WZW $\sgm$-model action is invariant under the $N=(2,2)$ supersymmetry transformations, these complex structures are also covariantly constant with respect to the corresponding connections with torsions. As has been shown in \cite{LyakhZ}, the bi-Hermitian (GK) geometry consistency condition for these connections is nothing else but the bi-Poisson structure on the $\sgm$-model target space.

\vskip 10 pt
\leftline{\bf 4.Bi-Poisson structure and GK geometry in Kazama-Suzuki models.}

 Here I consider the classical limit of the calculations from section 2 and show that Kazama-Suzuki conditions for the denominator subgroup $H$ determine in classical limit GK geometry on the target space of the corresponding N=2 supersymmetric $\sgm$-model.

 Our approach is based on the geometric description of $G/H$ coset models which is provided when they are considered as gauged WZW models \cite{GawK}, \cite{BaRS}. Integrating out the gauge fields we obtain $\sgm$-model action with some target space metric and $B$-field. In Hamiltonian approach \cite{AlSS} it causes the first class constraints for the currents valued in denominator subalgebra of the coset model. In the supersymmetric case at hand the constraints are 
\ber
L^{\al}(Z)-R^{\al}(Z)=0, \ L_{\al}(Z)-R_{\al}(Z)=0
\label{3.constr}
\enr
so that the observables $O(Z)$ of Kazama-Suzuki model are determined by the equation: 
\ber
\lbrc (L^{\mu}-R^{\mu})(Z_{1}),O(Z_{2})\rbrc=0
\label{3.fconstr}
\enr
where $L^{\mu}(Z)=(L^{\al}(Z),L_{\al}(Z))$, $R^{\mu}(Z)=(R^{\al}(Z),R_{\al}(Z))$. In case the denomonator subgroup contains direct abelian factors with trivial $Ad$-action the corresponding constraints from (\ref{3.constr}), (\ref{3.fconstr}) must be replaced by $L^{\mu}(Z)+R^{\mu}(Z)=0$. This circumstance will be implied hereinafter.

 Due to the discussion in section 3 it allows in particular to identify the functions on the target space of the Kazama-Suzuki $\sgm$-model with the $Ad_{H}$-invariant functions on $G$. So we can extend the 
disscussion of the section 3 to the case of Kazama-Suzuki coset model and show that Kazama-Suzuki conditions define bi-Poisson structure underlying the GK geometry in coset model. 

 First of all we use the classical Kazama-Suzuki coset model spin-1 currents
\ber
K^{L}_{KS}(Z)={\im\ov 2k}(\om_{Lij}L^{i}L^{j}(Z)-\om_{L\mu\nu}L^{\mu}L^{\nu}(Z))={1\ov k} L^{a}L_{a}(Z), 
\nmb
K^{R}_{KS}(Z)=-{\im\ov 2k}(\om_{Rij}R^{i}R^{j}(Z)-\om_{R\mu\nu}R^{\mu}R^{\nu}(Z))=-{1\ov k} R^{a}R_{a}(Z)
\label{3.KScarrent}
\enr
to define bilinear operations on the fields $\Psi_{1,2}(Z)$ satisfying the constraints (\ref{3.fconstr})
\ber
\lbrc \Psi_{1}(Z_{1}),\Psi_{2}(Z_{2})\rbrc_{K^{L}_{KS}}:=\lbrc K^{L}_{KS-1/2}\Psi_{1}(Z_{1}),\Psi_{2}(Z_{2})\rbrc,
\nmb
\lbrc \Psi_{1}(Z_{1}),\Psi_{2}(Z_{2})\rbrc_{K^{R}_{KS}}:=\lbrc K^{R}_{KS-1/2}\Psi_{1}(Z_{1}),\Psi_{2}(Z_{2})\rbrc
\label{3.KSDerbrack}
\enr 
In the case when the Kazama-Suzuki observables $\Psi_{1,2}(Z)$ are coming from the $Ad_{H}$-invariant functions $\Psi_{1,2}$ on $G$, these brackets define by the residues the skew-symmetric brackets on $\Psi_{1,2}$:
\ber
\lbrc \Psi_{1}(Z_{1}),\Psi_{2}(Z_{2})\rbrc_{K^{L}_{KS}}=
\dlt(Z_{1}-Z_{2}){\im\ov 2k}\om_{Lpq}(L^{p}\cdot\Psi_{1})(L^{q}\cdot\Psi_{2})(Z_{2})
\nmb
\lbrc \Psi_{1}(Z_{1}),\Psi_{2}(Z_{2})\rbrc_{K^{R}_{KS}}=
-\dlt(Z_{1}-Z_{2}){\im\ov 2k}\om_{Rpq}(R^{p}\cdot\Psi_{1})(R^{q}\cdot\Psi_{2})(Z_{2})
\label{3.KSDerbrack1}
\enr
where the common notation $L^{p}=(L^{a}(Z),L_{a}(Z))$, $R^{p}=(R^{a}(Z),R_{a}(Z))$ has been used.

 The next step is to show that due to (\ref{3.fconstr}) the algebra of $Ad_{H}$-invariant functions is closed under these brackets and the linear combinations
\ber
\lbrc \Psi_{1}(Z_{1}),\Psi_{2}(Z_{2})\rbrc_{\pm}:=
\lbrc \Psi_{1}(Z_{1}),\Psi_{2}(Z_{2})\rbrc_{K^{L}_{KS}}\pm \lbrc \Psi_{1}(Z_{1}),\Psi_{2}(Z_{2})\rbrc_{K^{R}_{KS}}
\label{3.KSDerbrack2}
\enr
define bi-Poisson structure on Kazama-Suzuki model target space by the residues.

 To this end one needs to calculate the Lie derivative of the brackets (\ref{3.KSDerbrack2}) with respect to
the denominator currents $L^{\mu}(Z)-R^{\mu}(Z)$. We obtain from the one hand
\ber
\lbrc L^{\mu}(W)-R^{\mu}(W),\lbrc \Psi_{1}(Z_{1}),\Psi_{2}(Z_{2})\rbrc_{\pm}\rbrc-
\lbrc\lbrc L^{\mu}(W)-R^{\mu}(W),\Psi_{1}(Z_{1})\rbrc,\Psi_{2}(Z_{2})\rbrc_{\pm}
\nmb
+\lbrc \Psi_{1}(Z_{1}),\lbrc  L^{\mu}(W)-R^{\mu}(W),\Psi_{2}(Z_{2})\rbrc\rbrc_{\pm}=
\nmb
\lbrc L^{\mu}(W)-R^{\mu}(W),\lbrc \Psi_{1}(Z_{1}),\Psi_{2}(Z_{2})\rbrc_{\pm}\rbrc
\label{3.Liederiv1}
\enr
From the other hand one can see from the definitions (\ref{3.KSDerbrack1}) that
\ber
\lbrc L^{\mu}(W)-R^{\mu}(W),\lbrc \Psi_{1}(Z_{1}),\Psi_{2}(Z_{2})\rbrc_{\pm}\rbrc-
\lbrc\lbrc L^{\mu}(W)-R^{\mu}(W),\Psi_{1}(Z_{1})\rbrc,\Psi_{2}(Z_{2})\rbrc_{\pm}
\nmb
+\lbrc \Psi_{1}(Z_{1}),\lbrc  L^{\mu}(W_)-R^{\mu}(W),\Psi_{2}(Z_{2})\rbrc\rbrc_{\pm}=
\nmb
\oint dZ\lbrc\lbrc\lbrc  L^{\mu}(W),K^{L}_{KS}(Z)\rbrc,\Psi_{1}(Z_{1})\rbrc,\Psi_{2}(Z_{2})\rbrc\pm
\nmb
\oint dZ\lbrc\lbrc\lbrc  R^{\mu}(W),K^{R}_{KS}(Z)\rbrc,\Psi_{1}(Z_{1})\rbrc,\Psi_{2}(Z_{2})\rbrc
\label{3.Liederiv2}
\enr
The result of Poisson brackets $\lbrc  L^{\mu}(W),K^{L}_{KS}(Z)\rbrc$, $\lbrc  R^{\mu}(W),K^{R}_{KS}(Z)\rbrc$ calculations can be read of from the
$(\cw-\cz)^{-{1\ov 2}}$ and $(\bar{\cw}-\bar{\cz})^{-{1\ov 2}}$ poles in the OPE's $L^{\mu}(\cw)K^{L}_{KS}(\cz)$ and $R^{\mu}(\bar{\cw})K^{R}_{KS}(\bar{\cz})$. But they vanish as we have seen in section 2. Thus we obtain
\ber
\lbrc L^{\mu}(W)-R^{\mu}(W),\lbrc \Psi_{1}(Z_{1}),\Psi_{2}(Z_{2})\rbrc_{\pm}\rbrc=0
\label{3.Liederiv3}
\enr
In view of (\ref{3.KSDerbrack1}) it means that algebra of $Ad_{H}$-invariant functions is closed
under the brackets which are deterined by the residues from the formula (\ref{3.KSDerbrack2}). Notice the close similarity of the reasoning here to the Poisson-homogeneous space reduction theorem from \cite{SemTian}.

 Now we show that the brackets (\ref{3.KSDerbrack2}) define bi-Poisson structure on invariant functions and
the Schouten brackets between the corresponding bivectors are proportional to the 3-form on the Kazama-Suzuki target space. Thus one needs to calculate two Jacobiators:
\ber 
\lbrc \Psi_{1}(Z_{1}),\lbrc \Psi_{2}(Z_{2}),\Psi_{3}(Z_{3})\rbrc_{K^{L}_{KS}}\rbrc_{K^{L}_{KS}}-
\lbrc \lbrc \Psi_{1}(Z_{1}),\Psi_{2}(Z_{2})\rbrc_{K^{L}_{KS}},\Psi_{3}(Z_{3})\rbrc_{K^{L}_{KS}}+
\nmb
\lbrc \Psi_{2}(Z_{2}),\lbrc \Psi_{1}(Z_{1}),\Psi_{3}(Z_{3})\rbrc_{K^{L}_{KS}}\rbrc_{K^{L}_{KS}}
\label{3.Jacob1}
\enr
\ber
\lbrc \Psi_{1}(Z_{1}),\lbrc \Psi_{2}(Z_{2}),\Psi_{3}(Z_{3})\rbrc_{K^{R}_{KS}}\rbrc_{K^{R}_{KS}}-
\lbrc \lbrc \Psi_{1}(Z_{1}),\Psi_{2}(Z_{2})\rbrc_{K^{R}_{KS}},\Psi_{3}(Z_{3})\rbrc_{K^{R}_{KS}}+
\nmb
\lbrc \Psi_{2}(Z_{2}),\lbrc \Psi_{1}(Z_{1}),\Psi_{3}(Z_{3})\rbrc_{K^{R}_{KS}}\rbrc_{K^{R}_{KS}}
\label{3.Jacob2}
\enr
They can be calculated directly, but the result is dictated by the 3-currents contributions in the OPE's $K^{L}_{KS}(\cz_{1})K^{L}_{KS}(\cz_{2})$ and
$K^{R}_{KS}(\bar{\cz}_{1})K^{R}_{KS}(\bar{\cz}_{2})$ so that the right hand sides of the Jacobiators (\ref{3.Jacob1}) and (\ref{3.Jacob2})
are equal correspondingly to
\ber
\dlt(Z_{2}-Z_{3})\dlt(Z_{1}-Z_{3}){1\ov 3k^{2}}(f_{ijk}(L^{i}\cdot\Psi_{1})(L^{j}\cdot\Psi_{2})(L^{k}\cdot\Psi_{3})(Z_{3})-
\nmb
f_{\nu\mu\lm}(L^{\nu}\cdot\Psi_{1})(L^{\mu}\cdot\Psi_{2})(L^{\lm}\cdot\Psi_{3})(Z_{3}))
\label{3.Jaco3}
\enr
and
\ber
-\dlt(Z_{2}-Z_{3})\dlt(Z_{1}-Z_{3}){1\ov 3k^{2}}(f_{ijk}(R^{i}\cdot\Psi_{1})(R^{j}\cdot\Psi_{2})(R^{k}\cdot\Psi_{3})(Z_{3})-
\nmb
f_{\nu\mu\lm}(R^{\nu}\cdot\Psi_{1})(R^{\mu}\cdot\Psi_{2})(R^{\lm}\cdot\Psi_{3})(Z_{3}))
\label{3.Jacob4}
\enr
Expressing the right translation action on functions in terms of the left translations and using (\ref{3.fconstr}) one can rewrite
(\ref{3.Jacob4}) as
\ber
-\dlt(Z_{2}-Z_{3})\dlt(Z_{1}-Z_{3}){1\ov 3k^{2}}(f_{ijk}(L^{i}\cdot\Psi_{1})(L^{j}\cdot\Psi_{2})(L^{k}\cdot\Psi_{3})(Z_{3})-
\nmb
f_{\nu\mu\lm}(L^{\nu}\cdot\Psi_{1})(L^{\mu}\cdot\Psi_{2})(L^{\lm}\cdot\Psi_{3})(Z_{3}))
\label{3.Jacob5}
\enr
Therefore the brackets (\ref{3.KSDerbrack2}) satisfy Jacobi super-identities. By the residues they define the pair of Poisson brackets on $Ad_{H}$-invariant functions on $G$ and the Schouten bracket of the bivector associated to $K^{L}_{KS}$ with the bivector associated to $K^{R}_{KS}$ is proportional to the 3-form on Kazama-Suzuki $\sgm$-model target space. In other words we get a bi-Poisson structure similarly to that we have found at the end of Section 3. The Poisson bivectors of this structure are given by a pair of complex structures which are skew-symmetric w.r.t. the $\sgm$-model target space metric and covariantly constant w.r.t. a pair of connections with torsion because of the $N=(2,2)$ Virasoro algebra is a symmetry algebra of the Kazama-Suzuki $\sgm$-model. This bi-Poisson structure is nothing else but the connections consistency condition \cite{LyakhZ} of the corresponding bi-Hermitian (GK) geometry.

 Thus, we proved that Kazama-Suzuki conditions
for the denominator subgroup $H$ of $N=2$ superconformal $G/H$ coset model determine GK geometry on the target space of the corresponding $\sgm$-model. 

\vskip 10pt
\leftline{\bf 5. Conclusion.}

 The $N=2$ supersymmetric models of Conformal Field Theory play an important role in construction of realistic models of superstring in 4 dimensions. In this paper we analysed the Kazama-Suzuki coset construction of unitary $N=2$ superconformal field theory models in terms of Generalized K$\ddot{a}$hler geometry. It is swown that conditions for the denominator subgroup of N=2 superconformal $G/H$ coset model found by Kazama and Suzuki determine Generalized K$\ddot{a}$hler geometry on the target space of the corresponding N=2 supersymmetric $\sgm$-model. So, the Kazama-Suzuki coset construction can be considerd as a tool to produce new examples of GK geometry $\sgm$-model target spaces whose quantum field theories are known.

 As it has already been mentioned the arguments presented during our analysis are close to the Poisson-homogeneous space reduction theorem, so it would be interesting to see whether the Kazama-Suzuki construction can be undersood as a generalized complex geometry reduction developed in \cite{SHu1}, \cite{SHu2}.

 The Kazama-Suzuki coset models exhibit also the $W$-superalgebras of conserved currents \cite{Ahn} which makes them exactly solvable. It would be important to investigate these algebras from the point of view of GK geometry.


 Another question to study is to understand geometricaly the Gepner's construction of superstring compactification. It is interesting to see whether GK geometry (or rather its quantum version) can be used to reproduce the Hodge numbers of Calabi-Yau $\sgm$-model compactification.

\vskip 10pt
\centerline{\bf ACKNOWLEDGEMENTS}
\frenchspacing
 The work has been supported by the Russian Science Foundation under the grant 18-12-00439.

\end{document}